\documentclass[12pt,letterpaper]{article}

\usepackage{amsmath}
\usepackage{amsfonts}
\usepackage{amssymb}
\usepackage{graphicx}
\usepackage[numbers,sort&compress]{natbib}

\usepackage[hang,small,bf]{caption}

\DeclareRobustCommand{\SkipTocEntry}[4]{}

\def\beq{\begin{equation}}
\def\eeq{\end{equation}}
\def\bea{\begin{eqnarray}}
\def\eea{\end{eqnarray}}

\begin{document}

\vspace{5mm} \vspace{0.5cm}

\begin{center}

{\large {CMBPol} Mission Concept Study} \\
	\vskip 15pt {\Large {CMBPol}: A Mission to Map our Origins}
	\\[1.0cm]

{ Daniel Baumann$^{\rm 1,2,3}$, Asantha Cooray$^{\rm 4}$, Scott Dodelson$^{\dagger}$$^{\rm 5,6,7}$, 
Joanna Dunkley$^{\rm 8,3}$,\\
Aur\'elien~A.~Fraisse$^{\rm 9}$, Mark G.~Jackson$^{\rm 5, 10,11}$, Al Kogut$^{\rm 12}$, Lawrence M.~Krauss$^{\rm 13}$, \\ 
Kendrick M. Smith$^{\rm 14}$, and Matias Zaldarriaga$^{\rm 1,2}$}
\\[1.5cm]

\end{center}

\begin{center}


{\small \textit{$^{\rm 1}$ Department of Physics, Harvard University, Cambridge, MA 02138, USA}}

{\small \textit{$^{\rm 2}$ Center for Astrophysics, Harvard University, Cambridge, MA 02138, USA}}

{\small \textit{$^{\rm 3}$ Department of Physics, Princeton University, Princeton, NJ 08540, USA}}

{\small \textit{$^{\rm 4}$ Center for Cosmology, University of California, Irvine, CA 92697, USA}}

{\small \textit{$^{\rm 5}$ Center for Particle Astrophysics, Fermilab, Batavia, IL 60510, USA}}

{\small \textit{$^{\rm 6}$ Department of Astronomy and Astrophysics, University of Chicago, Chicago, IL 60637, USA}}

{\small \textit{$^{\rm 7}$ Kavli Institute for Cosmological Physics, Chicago, IL 60637, USA}}

{\small \textit{$^{\rm 8}$ Astrophysics Department, University of Oxford, Oxford, OX1 3RH, UK}}

{\small \textit{$^{\rm 9}$ Princeton University Observatory, Peyton Hall, Princeton, NJ 08544, USA}}

{\small \textit{$^{\rm 10}$ Theory Group, Fermilab, Batavia, IL 60510, USA}}

{\small \textit{$^{\rm 11}$ Lorentz Institute for Theoretical Physics, 2333CA Leiden, the Netherlands}}

{\small \textit{$^{\rm 12}$ NASA/Goddard Space Flight Center, Greenbelt, MD 20771, USA}}

{\small \textit{$^{\rm13}$ School of Earth and Space Exploration, Arizona State University, Tempe, AZ 85287, USA}}

{\small \textit{$^{\rm 14}$ Institute of Astronomy, University of Cambridge, Cambridge, CB3 0HA, UK}}

\end{center}

\vspace{1.5cm} \hrule \vspace{0.3cm}
{\small  \noindent \textbf{Abstract} \\[0.3cm]
\noindent 
Quantum mechanical metric fluctuations during an early inflationary phase of the universe
 leave a characteristic imprint in the 
polarization of the cosmic microwave background (CMB). The amplitude of this signal depends on the energy scale at which inflation occurred. Detailed observations by a dedicated satellite mission
({\sl CMBPol}) therefore provide information about energy scales as high as $10^{15}$~GeV, twelve orders of magnitude greater than the highest energies accessible to particle accelerators, and probe the earliest moments in the history of the universe. This summary provides an overview of
a set of studies exploring the scientific payoff of {\sl CMBPol} in diverse areas of modern cosmology, such as the physics of inflation \cite{BaumannInflation}, gravitational lensing \cite{SmithLensing} and cosmic reionization \cite{ZaldarriagaReionization}, as well as foreground science \cite{FraisseFGs} and removal \cite{DunkleyFGs}.
 \vspace{0.5cm}  \hrule
\def\thefootnote{\arabic{footnote}}
\setcounter{footnote}{0}

\vspace{1.0cm}

\vfill \noindent
$^\dagger$ {\footnotesize {\tt dodelson@fnal.gov}}\\



Dramatic new observations made over the past decade have led to a new standard model of  cosmology: a flat universe dominated by unknown forms of dark energy and dark matter.  Beyond these general features, observations of primordial perturbations in the cosmic microwave background (CMB) also contain clues about the physics of the early universe that may help us resolve outstanding puzzles of particle physics and cosmology.  

This new standard cosmological model is based on the idea that early in its history, the universe underwent a brief epoch of {\it inflation} during which the expansion accelerated and small-scale fluctuations were stretched to superhorizon scales. The theory posits that quantum mechanical fluctuations in the fields that describe the matter content at early times seeded the structure in the universe, leading to two important signatures:
\begin{itemize} 
  \item Primordial anisotropies imprinted in the CMB at a level of a few parts in a hundred thousand. 
  \item An inhomogeneous late time matter distribution in the form of a  cosmic web as traced for example by the location of galaxies in the universe. 
  \end{itemize}  
  The inflationary paradigm successfully predicted the detailed statistical features of both of these signatures.
 However, the fundamental microphysics behind inflationary models is difficult to probe directly because inflation is likely to involve energy scales as high as $10^{15}$~GeV, well beyond the energies which can be accessed at accelerators. The inferential nature of the evidence to date allows not only a plethora of underlying models for the inflationary expansion but perhaps even alternatives which do not call for accelerated expansion. 

There is one prediction of inflation, however, that can resolve many of the current ambiguities. 
Anisotropies in the CMB and the cosmic web of galaxies are seeded by {\it scalar} perturbations during inflation, which produce spatial variations in the Newtonian gravitational potential and the density of matter.  However, scalar perturbations are not the only perturbations that are generated during inflation.  If the energy scale of inflation is high enough, quantum fluctuations in the gravitational field during inflation result in an observable stochastic background of gravitational waves, or {\it tensor} modes. 

While such a primordial gravitational wave background exists on all scales, the greatest sensitivity to such a background comes from the detailed pattern of polarization in the CMB.  Thomson scattering of an anisotropic radiation background off of free electrons before the time of electron-proton recombination (when the universe was 380,000 years old) produced a radiation field polarized at the 10\% level. Since polarization is described at every position by an amplitude and an angle of orientation, the polarization field on the sky can be decomposed into two modes, a curl-free $E$-mode and a divergence-less $B$-mode, as depicted in Figure~\ref{fig:EBmode}. Crucially, the $B$-mode pattern cannot be produced by scalar perturbations.  {\bf The detection of a non-zero $B$-mode would therefore be a smoking-gun signature of primordial gravitational waves such as those predicted by inflation.}

\begin{figure}[htbp!]
    \centering
        \includegraphics[width=.45\textwidth]{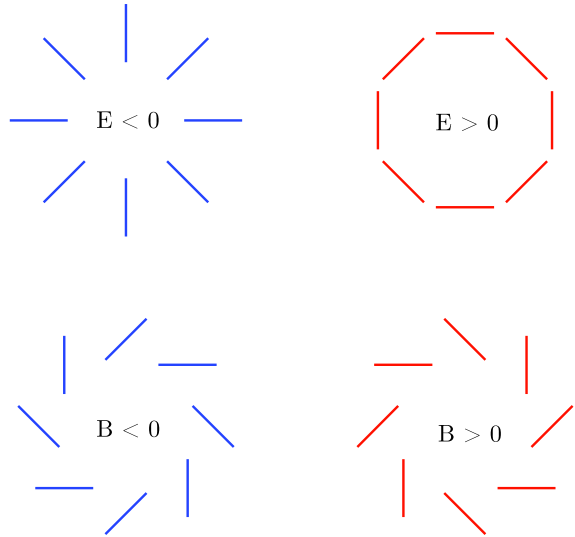}
   \caption{Examples of $E$-mode and $B$-mode patterns of polarization. 
   $E$-polarization is characterized by polarization vectors that are radial around cold spots and tangential around hot spots on the sky.
In contrast, $B$-polarization has a {\it curl}; i.e., its polarization vectors have vorticity around any given point on the sky.
Although $E$ and $B$ are both invariant under rotations, they behave differently under parity transformations.   
If reflected across a line going through the center the $E$-patterns are unchanged, while the positive and negative $B$-patterns get interchanged.}
    \label{fig:EBmode}
\end{figure}

\begin{figure}[h!]
    \centering
        \includegraphics[width=0.65\textwidth]{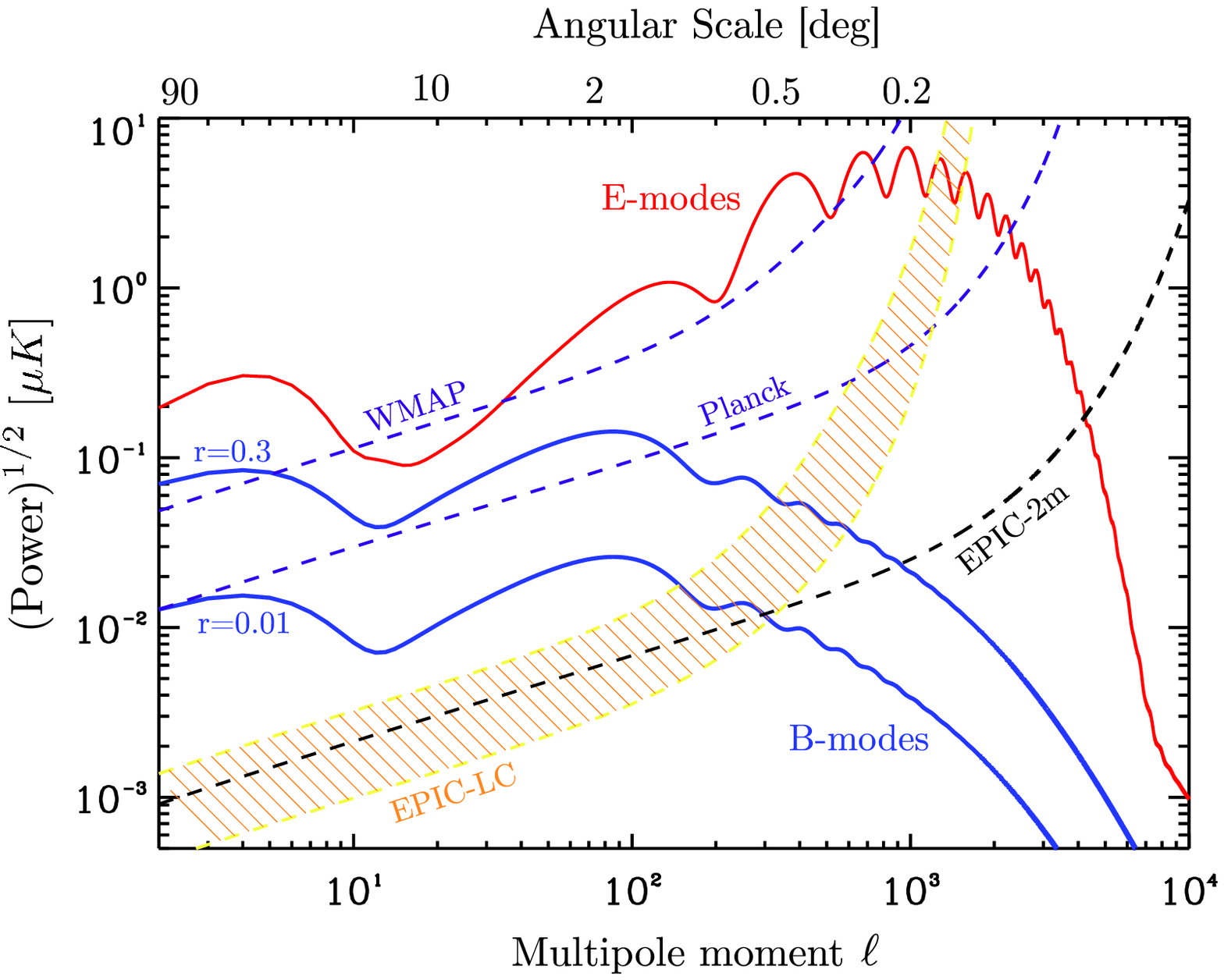}
    \caption{$E$- and $B$-mode power spectra for a tensor-to-scalar ratio saturating current bounds, $r=0.3$, and for $r=0.01$. Shown are also the experimental sensitivities for {\sl WMAP}, {\sl Planck} and two different realizations of {\sl CMBPol} (EPIC-LC and EPIC-2m). On large scales where much of the $B$-mode signal resides, {\sl CMBPol} is more than an order of magnitude more sensitive than {\sl Planck} or {\sl WMAP}. ({\small Figure adapted from Ref.~\cite{Bock:2008ww}.})}
    \label{fig:spectra}
\end{figure}

Several experiments have already detected the $E$-mode of polarization at the expected 10\% level and have begun measuring its spectrum (Figure~\ref{fig:spectra}). However, current measurements do not yet have high enough sensitivity to detect the $B$-modes which are expected to have a lower amplitude\footnote{The $B$-mode amplitude is determined by the ratio of power in tensor modes to the power in scalar modes.  This tensor-to-scalar ratio is denoted by the parameter $r$.}. Over the coming decade, the {\sl Planck} satellite and sub-orbital experiments will continue mapping the polarization of the CMB, will put limits on $B$-modes, and may even detect a signal. However, in order to fully explore the microphysics that could generate $B$-modes while at the same time providing a rich data set that can impact upon other important astrophysical investigations, a satellite mission dedicated to mapping polarization with high sensitivity and broad frequency coverage will be required.

Acknowledging these exciting possibilities, NASA funded the `CMBPol Mission Concept Study'. Part of this study was a theoretical effort centered around a workshop held in June 2008 at Fermilab. The workshop was attended by over 100 scientists with interests ranging from bolometers to string theory. Five working groups formed, each with the goal of developing a different part of the science case for {\sl CMBPol}. Since the workshop, dozens of scientists have contributed to these working groups in an effort to solidify this set of arguments. In a series of papers \cite{BaumannInflation, SmithLensing, ZaldarriagaReionization, FraisseFGs, DunkleyFGs} accompanying this summary, we present the anticipated benefits and challenges facing {\sl CMBPol}. The main conclusions of the five working groups are:

\begin{itemize}
\item {\it {\sl CMBPol} has the potential to vastly increase our understanding of the primordial mechanism responsible for the generation of structure in the universe} \cite{BaumannInflation}. 
Many models of inflation predict a gravitational wave signal that can be convincingly detected only with {\sl CMBPol} (see Figure \ref{fig:nsr}; $r > 0.01$). The detection of such a signal would advance our knowledge of fundamental physics at energy scales that cannot be probed by terrestrial accelerators. Even a  null result would have important physical implications as it would rule out a wide range of inflationary models, while leaving a well-defined class of inflationary mechanisms to consider.

\begin{figure}[h!]
    \centering
        \includegraphics[width=.70\textwidth]{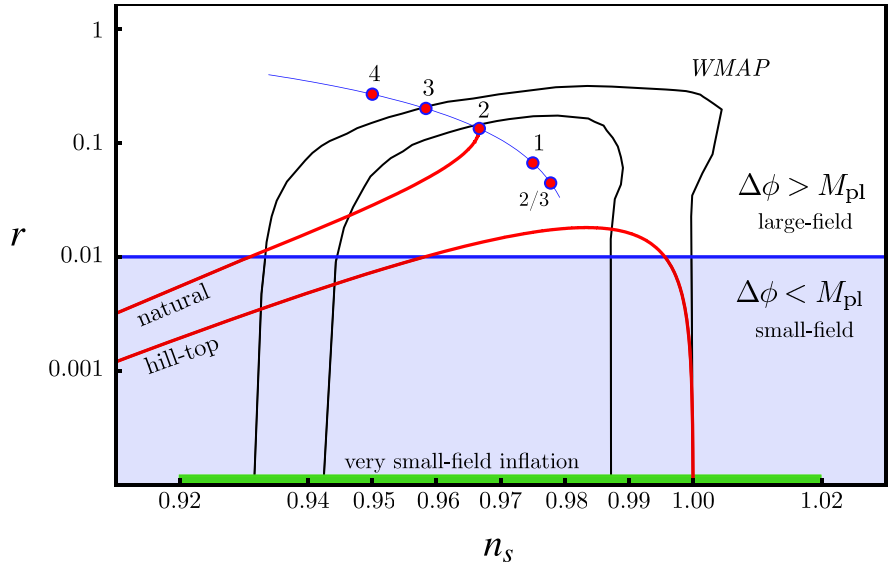}
   \caption{Constraints on single-field slow-roll models in the $n_s$-$r$ plane. The value of $n_s$ measures the scale-dependence of the scalar spectrum, while $r$ characterizes the amplitude of the tensor signal (and hence the energy scale of inflation). The value of $r$ furthermore determines whether the models involve large or small field variations during inflation.     
   Shown are the 5-year {\sl WMAP} constraints on $n_s$ and $r$ (area above the curves ruled out at 68 and 95\% confidence level) as well as
the predictions of a few representative models of single-field slow-roll inflation (colored lines; dots correspond to a power law potential $\phi^p$for the field driving inflation with $p=4, 3, 2, 1, \frac{2}{3}$). Generally, models in which this field changes considerably in Planck units during inflation predict values of $r$ greater than 0.01.
Realistic forecasts for {\sl CMBPol} specifications show that gravitational waves can be detected at $\sim 3 \sigma$ for $r>0.01$ \cite{BaumannInflation}. This shows that {\sl CMBPol} is a powerful instrument to test this crucial regime of the inflationary parameter space.}
    \label{fig:nsr}
\end{figure}

In addition, {\it observations made by {\sl CMBPol} will provide important new constraints on the non-Gaussianity of scalar fluctuations, on primordial isocurvature fluctuations, and on the scale-dependence of scalar and tensor modes.}  This provides crucial information about the detailed dynamics underlying the inflationary era.

Finally, {\sl CMBPol} measurements could be used to probe non-zero spatial curvature, large-scale anisotropy, or topological defects, any one of which would point to exotic physics. 

\item {\it The polarization pattern is observed through the gravitational lens of the inhomogeneous universe. Lensing contaminates the $B$-mode signal but also encodes important information about our universe} \cite{SmithLensing}. Photons traveling from the last scattering surface at redshift $z\simeq1090$ are deflected due to the spatially varying gravitational potentials. Lensing can therefore distort the canonical scalar-produced $E$-modes so that $B$-modes appear on small scales, and is thus a source of noise when attempting to extract a primordial gravitational wave signal. However, detailed analysis of the small-scale polarization field will also reveal information about the projected gravitational potential. This new information can be used not only to clean the noise due to lensing but also to constrain fundamental physics such as dark energy and neutrino masses. 

\item {\it Foregrounds can be adequately cleaned as long as {\sl CMBPol} has a wide range of frequency channels} \cite{DunkleyFGs}. Current measurements such as {\sl WMAP} already must model and subtract
contamination from galactic foregrounds at large angular scales. To detect the background of gravitational waves predicted by many inflationary models, {\sl CMBPol} must demonstrate a capacity to clean foreground contamination. Using state-of-the-art models for polarized foregrounds and modern methods of analysis, we demonstrate that {\sl CMBPol} will be able to extract even a small gravitational wave signal at the $1\%$ level ($r=0.01$).  

\item {\it {\sl CMBPol} will improve our knowledge of the magnetized Galactic interstellar medium, as well as of the properties of interstellar dust in our galaxy}~\cite{FraisseFGs}.  In particular, high sensitivity channels at low CMB frequencies will probe the sky-projected component of the Galactic magnetic field at intermediate and high Galactic latitudes over the whole sky for the first time.  These observations will provide information on the regular, turbulent and halo components of the field.  In addition, data provided by {\sl CMBPol}'s high frequency channels will be used to constrain models of interstellar dust, and observations in many patches of the sky at high Galactic latitudes will shed new light on the composition and alignment mechanism(s) of interstellar dust grains.  Finally, {\it our understanding of extra-Galactic source counts and spectral energy distributions will be tested} as we attempt to eliminate their contribution to CMB lensing potential estimators.

\item {\it Large-scale polarization data will enhance our understanding of the end of the dark ages and the formation of the first stars} \cite{ZaldarriagaReionization}. Reionization at $z\simeq10$ produces a distinctive signature in the CMB polarization field on large scales. {\sl WMAP} has already used its maps to constrain the epoch of reionization and {\sl Planck} will improve on these constraints. {\sl CMBPol} will go even further: its cosmic variance-limited large angle maps will contain virtually all the information encoded in the CMB about reionization.

\end{itemize}

{\noindent \bf Conclusions:} 
A satellite mission to measure CMB polarization with unprecedented sensitivity and
frequency coverage could propel our empirical understanding of fundamental physics and cosmology by orders of magnitude in ways that are otherwise inaccessible to experiment, and could help answer many of the outstanding mysteries surrounding the seeds of structure in the universe. Even if {\sl CMBPol}  detects neither $B$-modes, non-Gaussianities in the primordial spectrum, nor isocurvature or non-zero spatial curvature or topological defects (all of which will be probed very sensitively by {\sl CMBPol}), the mission
would still deliver immense scientific payoff.  A non-detection of $B$-modes would rule out the most
widely studied class of inflationary models, thereby redirecting particle physics model building in general and inflationary theories in particular. Even in this pessimistic scenario, {\sl CMBPol} would still 
measure the projected gravitational potential weighted at redshifts much larger than those probed in galaxy surveys thereby
constraining dark energy and neutrino masses.  It would also extract all information encoded in the CMB about reionization,
map out magnetic fields and dust in our Galaxy, and add important information about extra-Galactic
sources. On the other hand, a detection of primordial $B$-modes would provide one of the most important observations ever made in cosmology and would extend our window on the universe back to the earliest moments of the Big Bang.
\medskip

{\noindent \bf Acknowledgments:} 
This research was partly funded by NASA Mission Concept Study award NNX08AT71G S01. We also acknowledge the organizational work of the Primordial Polarization Program Definition Team.




\end{document}